\documentclass[preprint,review,12pt]{elsarticle}

\usepackage{amsmath}
\usepackage{amsfonts}

\usepackage{caption}
\usepackage{subcaption}

\usepackage[colorlinks=true]{hyperref}

\journal{Computers and Mathematics with Applications}

\biboptions{sort&compress}

\begin{document}

\begin{frontmatter}

\title{Evaluation of the Finite Element Lattice Boltzmann Method for Binary Fluid Flows}

\author[nbi] {Rastin Matin\corref{cor1}}
\ead{rastin@nbi.ku.dk}
\author[nbi] {Marek Krzysztof Misztal\corref{cor1}}
\ead{misztal@nbi.ku.dk}
\author[nbi] {Anier Hern\'{a}ndez-Garc\'{i}a}
\ead{ahernan@nbi.ku.dk}
\author[nbi] {Joachim Mathiesen}
\ead{mathies@nbi.ku.dk}

\cortext[cor1]{Corresponding authors}

\address[nbi]  {Niels Bohr Institute, University of Copenhagen, DK-2100 Copenhagen, Denmark}

\begin{abstract}
In contrast to the commonly used lattice Boltzmann method, off-lattice Boltzmann methods decouple the velocity discretization from the underlying spatial grid, thus allowing for more efficient geometric representations of complex boundaries. The current work combines characteristic-based integration of the streaming step with the free-energy based multiphase model by Lee \emph{et. al.} [\emph{Journal of Computational Physics, 206 (1), 2005}].
This allows for simulation time steps more than an order of magnitude larger than the relaxation time. Unlike previous work by Wardle \emph{et. al.} [\emph{Computers and Mathematics with Applications, 65 (2), 2013}] 
that integrated intermolecular forcing terms in the advection term, the current scheme applies collision and forcing terms locally for a simpler finite element formulation. A series of thorough benchmark studies reveal that this does not compromise stability and that the scheme is able to accurately simulate flows at large density and viscosity contrasts. 
\end{abstract}

\begin{keyword}
Lattice Boltzmann method \sep Finite element method \sep Multiphase flows
\MSC[2010] 0898-1221
\end{keyword}

\end{frontmatter}

\section{Introduction}
The lattice Boltzmann method has become popular as a numerical solver for multiphase flows. Several models have been proposed in the literature during the last two decades that can generally be classified in four categories: The chromodynamic model by Gunstesen \emph{et. al.} \cite{PhysRevA.43.4320, 0295-5075-18-2-012} which was used for the earliest simulations, the phenomenological interparticle-potential model by Shan and Chen \cite{PhysRevE.47.1815, PhysRevE.49.2941},
the free-energy model by Swift \emph{et. al.} \cite{PhysRevLett.75.830} and
the mean-field model by He \emph{et. al.} \cite{He1999642, PhysRevE.57.R13} based on the kinetic theory for dense fluids. With the exception of the latter, these models are restricted to Boussinesq flows in their original formulations, a limit which has been pushed by the many recent developments described in several reviews, see \emph{e.g.} \cite{MP_Huang, JHarting, Lee2012Connington}.


\subsection{Overview}
Simulating binary flows beyond the Boussinesq approximation is generally a challenging issue due to the sharp changes in density across the interface. 
In an attempt to remedy this, He \emph{et. al.} \cite{He1999642} introduced an incompressible transformation in their kinetic model, changing the particle distribution function for mass and momentum into that for hydrodynamic pressure and momentum. 
Adding to this transformation, Lee and Lin \cite{Lee200516} enhanced stability of their free-energy based model by adopting the stress form of the surface tension force for the pressure-momentum lattice Boltzmann equation (LBE) and the potential form of the surface tension force for the LBE of the order parameter. They furthermore introduced discretization schemes that comply with the second-order accuracy of the lattice Boltzmann method, and their model has recently been augmented to allow arbitrary wetting properties of the two phases \cite{Lee20108045}. All of the above models have been shown to produce stable and accurate results for density ratios of up to 1000 and viscosity ratios up to 50. Here, we shall consider further the model of Lee and Lin \cite{Lee200516}.

\subsection{Off-Lattice Boltzmann Methods}
The traditional, regular-grid based setting limits the application of the lattice Boltzmann method to uniform Cartesian grids. However, extensions have been made to irregular grids by introducing a class of off-lattice Boltzmann schemes consisting of finite volume \cite{FLD:FLD1018, Patil20095262} and finite element schemes \cite{Lee2001336, 0295-5075-75-3-434}. Inherent to all of these is the standard Courant-Friedrichs-Lewy (CFL) condition on the time step $\delta t$, a necessary condition for the stability of any kind of advection equation. Certain schemes employ an explicit treatment of the collision term, thereby imposing the more restrictive condition $\delta t < 2\tau$ for forward Euler time integration and $\delta t < \tau$ for Strang splitting, where $\tau$ is the relaxation time \cite{Succi2008, Misztal2015316, Dellar:2013:IDL:2425424.2425484}. 


Characteristic-based schemes generally tend to provide better numerical stability compared to other time-integration schemes such as Runge-Kutta \cite{Rao2015251} and furthermore allow for an implicit integration of the collision term. This property is utilized in \cite{Lee2001336} to allow for CFL numbers up to 100 at the expense of increased computation time per time step by employing an iterative solver for the implicit term. In more recent work a variable transformation is often employed that masks the implicitness while preserving mass and momentum conservation \cite{0295-5075-75-3-434, PhysRevE.67.066709, He1998282}. Combining this with explicit second-order accurate Crank-Nicolson time integration, Bardow \emph{et. al.} \cite{0295-5075-75-3-434} successfully overcome the restrictive collision time step condition. 

However, this variable transformation does not preserve mass and momentum for the present multiphase model due to the form of the intermolecular forcing term describing fluid-fluid interaction. The current work instead applies the BGK-collision and forcing locally in the collision step, which allows for time steps more than an order of magnitude larger than the relaxation time when combined with second-order accurate advection.

\section{Numerical Method}
\subsection{Lattice Boltzmann Method}
The current study uses the model initially presented in \cite{Lee200516} in three dimensions, which introduces two particle distribution functions $f_\alpha$ and $g_\alpha$. The distribution function $f_\alpha$ recovers the order parameter (density) that tracks the interface between the two different phases and $g_\alpha$ recovers the hydrodynamic flow fields (pressure and momentum) of the two fluids. As the two distribution functions have different purposes, the stress and potential forms of the surface tension force are selectively adopted to match their roles. 

Integrating the governing discrete Boltzmann equation for $f_\alpha$ and $g_\alpha$ over a time step $\delta t$ and applying the trapezoidal rule leads to (\cite{Lee200516})
\begin{align}
f_\alpha(x_i+e_{\alpha i}\delta t, t+\delta t) - f_\alpha(x_i, t) = {}& + (-\Omega_{f_\alpha} + F_\alpha)|_{(x_i, t)} \label{eqn::lbmf}       \\
                                                                      & + (-\Omega_{f_\alpha} + F_\alpha)|_{(x_i + e_{\alpha i} \delta t, t+\delta t)} \notag \\
g_\alpha(x_i+e_{\alpha i}\delta t, t+\delta t) - g_\alpha(x_i, t) = {}& + (-\Omega_{g_\alpha} + G_\alpha + \mathcal G_\alpha)|_{(x_i, t)} \label{eqn::lbmg}    \\
                                                                      & + (-\Omega_{g_\alpha} + G_\alpha + \mathcal G_\alpha)|_{(x_i + e_{\alpha i} \delta t, t+\delta t)} \notag
\end{align}
where the intermolecular forcing terms $F_\alpha$ and $G_\alpha$ and the BGK-operator $\Omega_{\psi_\alpha}$ for a given distribution function $\psi \in \{f,g\}$ are given by
\begin{align}
\Omega_{\psi_\alpha} =& + \frac{1}{2\tau}(\psi_\alpha - \psi_\alpha^{\text{eq}}) \label{eq:omega} \\ 
F_\alpha             =& + \frac{\delta t}{2} \frac{(e_{\alpha i}-u_i)[\partial_i \rho c_s^2-\rho \partial_i(\mu_\varphi - \kappa \partial_k\partial_k\rho)]}{c_s^2}\Gamma_\alpha(u_i) \label{eq:F} \\
G_\alpha           = &+ \frac{\delta t}{2} \frac{(e_{\alpha i}-u_i)\partial_i \rho c_s^2}{c_s^2}[\Gamma_\alpha(u_i)-\Gamma_\alpha(0)] \label{eq:G} \\
			          & + \frac{\delta t}{2} \frac{(e_{\alpha i}-u_i)[\kappa\partial_i(\partial_k \rho \partial_k \rho) - \kappa \partial_j (\partial_i \rho \partial_j \rho)]}{c_s^2}\Gamma_\alpha(u_i) \notag
\end{align}
and $\mathcal G_\alpha$ is a volumetric body force. Here $e_{\alpha i}$ denote the 19 discrete particle velocities in directions $\alpha$ of the D3Q19 model, $c_s=1/\sqrt{3}$ the constant speed of sound, $u_i$ the macroscopic velocity, $\rho$ the mixture density and $\mu_\varphi$ the chemical potential. The dimensionless relaxation parameter $\tau$ is proportional to the kinematic viscosity $\nu$ through $\nu=c_s^2 \tau \delta t $. The equilibrium distribution functions $f_\alpha^{\text{eq}}$ and $g_\alpha^{\text{eq}}$ are given by
\begin{align}
f_\alpha^{\text{eq}} &= w_\alpha \rho \bigg[1+ \frac{e_{\alpha i}u_i}{c_s^2} + \frac{(e_{\alpha i}e_{\alpha j}-c_s^2\delta_{ij})u_iu_j}{2c_s^4}\bigg] \label{eqn::f_eq}\\
g_\alpha^{\text{eq}} &= w_\alpha  \bigg[\frac{p}{c_s^2}+ \rho\bigg(\frac{e_{\alpha i}u_i}{c_s^2} + \frac{(e_{\alpha i}e_{\alpha j}-c_s^2\delta_{ij})u_iu_j}{2c_s^4}\bigg)\bigg] \label{eqn::g_eq}
\end{align}
and $\Gamma(u_i) = f_\alpha^{\text{eq}}/\rho$, where $w_\alpha$ is the integral weighting factors of the D3Q19 model. The constants $\beta$ and $\kappa$ are determined by the surface tension $\sigma$ and interface width $\xi$
\begin{align}
\beta = \frac{12\sigma}{\xi(\rho_h-\rho_l)^4}, \qquad \kappa = \frac{3}{2}\frac{\xi\sigma}{(\rho_h-\rho_l)^2},
\end{align}
from which the chemical potential is explicitly given as 
\begin{align}
\mu_\varphi = 4\beta (\rho - \rho_v)(\rho - \rho_l)(\rho - 0.5(\rho_v+\rho_l)),
\end{align}
where $\rho_l$ $(\rho_v)$ denotes the bulk density of the liquid (vapor) phase. The relaxation parameter is given as the harmonic mean of the respective bulk relaxation parameters $(\tau_l , \tau_v)$ of the two phases weighted by the composition $C$ \cite{Lee20108045},
\begin{align}
\frac{1}{\tau} = \frac{C}{\tau_l} + \frac{1-C}{\tau_v},
\end{align}
where $C= (\rho-\rho_v)/(\rho_l-\rho_v)$.

\subsection{Numerical Scheme}
In order to solve the implicit equations \eqref{eqn::lbmf}-\eqref{eqn::lbmg} we first follow the procedure shown in \cite{Lee200516}, which splits the equations into the pre-streaming collision, streaming and post-streaming collision steps as follows
\paragraph{Pre-streaming collision}
\begin{align}
\bar{f}_\alpha(x_i, t) &= f_\alpha(x_i, t) + \bigg(\!\!-\Omega_{f_\alpha} + F_\alpha\bigg)\bigg|_{(x_i, t)} \label{eqn::lbmf_pre} \\
\bar{g}_\alpha(x_i, t) &= g_\alpha(x_i, t) + \bigg(\!\!-\Omega_{g_\alpha} + G_\alpha + \mathcal{G}_\alpha\bigg)\bigg|_{(x_i, t)} \label{eqn::lbmg_pre}
\end{align}
\paragraph{Streaming}
\begin{align}
\bar{f}_\alpha(x_i+e_{\alpha i}\delta t, t+\delta t) &= \bar{f}_\alpha(x_i, t) \label{eqn::lbmf_stream}       \\                                                                \bar{g}_\alpha(x_i+e_{\alpha i}\delta t, t+\delta t) &= \bar{g}_\alpha(x_i, t) \label{eqn::lbmg_stream}
\end{align}      
\paragraph{Post-streaming collision}
\begin{align}
f_\alpha(x_i+e_{\alpha i}\delta t, t+\delta t) ={}& + \bar{f}_\alpha(x_i+e_{\alpha i}\delta t, t+\delta t)\label{eqn::lbmf_post} \\ 
& + \frac{2\tau}{2\tau+1} \bigg(\!\!-\Omega_{f_\alpha} + F_\alpha\bigg)\bigg|_{(x_i + e_{\alpha i} \delta t, t+\delta t)} \notag \\
g_\alpha(x_i+e_{\alpha i}\delta t, t+\delta t) ={}& + \bar{g}_\alpha(x_i+e_{\alpha i}\delta t, t+\delta t)\label{eqn::lbmg_post} \\ 
& + \frac{2\tau}{2\tau+1}\bigg(\!\!-\Omega_{g_\alpha} + G_\alpha + \mathcal{G}_\alpha \bigg)\bigg|_{(x_i + e_{\alpha i} \delta t, t+\delta t)} \notag  
\end{align}
The volumetric body force is applied using the exact difference method \cite{Kupershtokh2009965}, 
\begin{align}
\mathcal G_\alpha = \frac{\delta t}{2}(g^{\text{eq}}_\alpha(\rho, u_i + \delta u_i) - g^{\text{eq}}_\alpha (\rho, u_i)),
\end{align}
where $\delta u_i=g_i\delta t$ for the case of gravity $g_i$. The density, momentum and hydrodynamic pressure are calculated by taking the zeroth and the first moments of the streamed distribution functions
\begin{align}
\rho     &= \sum_\alpha f_\alpha \\
\rho u_i &= \sum_\alpha e_{i\alpha} g_\alpha + \frac{\delta t}{2}\kappa \bigg[\partial_i(\partial_k \rho \partial_k \rho) - \partial_j (\partial_i \rho \partial_j \rho) \bigg] + \frac{\delta t}{2}\rho g_i \label{eq:velocity} \\
p        &= c_s^2\sum_\alpha g_\alpha + \frac{\delta t}{2}u_i \partial_i \rho c_s^2 \label{eq:pressure}
\end{align}

As both collision steps are performed locally, they do not require further work in order to be incorporated into an unstructured grid-based solver. The only term requiring further discretization is the streaming step, which will be described in the following.
\subsubsection{Finite Element Streaming}
In off-lattice schemes streaming is performed in an Eulerian sense, and the current work follows that of \cite{0295-5075-75-3-434, Zienkiewicz95, 10.3389/fphy.2015.00050} by applying Taylor expansion around $(x_i+e_{\alpha i}\delta t, t+\delta t)$ to Eqs. \eqref{eqn::lbmf_stream}-\eqref{eqn::lbmg_stream} in order to integrate them numerically
\begin{align}
\psi_\alpha^{n+1} &= \psi_\alpha^n - \delta t e_{\alpha i}\partial_i \psi_\alpha^n + \frac{\delta t^2}{2}e_{\alpha i}e_{\alpha j}\partial_i \partial_j \psi_\alpha^n  + \mathcal{O}(\delta t^3). \label{eq:local_formulation}
\end{align}
The streaming step \eqref{eq:local_formulation} is formally equivalent to that of Wardle \emph{et. al.} \cite{Wardle2013230, Lee2003445}. In contrast to the current work, the fluid-fluid interaction there is integrated in the streaming step and details concerning the nonlinear force term discretization are omitted. 

Eq. \eqref{eq:local_formulation} can now be discretized in space using the Galerkin finite element method, where spatial decomposition using linear, tetrahedral elements has been applied. The particle distribution functions are sampled at the vertices of the tetrahedral mesh and interpolated at other points,
\begin{align}
\psi_\alpha^n(\mathbf{x}) \approx \tilde{\psi}_\alpha^n(\mathbf{x}) = \mathbf{N}(\mathbf{x})^T \tilde{\boldsymbol{\psi}}_\alpha^n, \label{eq:discrete_field}
\end{align} 
where $\tilde{\psi}_\alpha^n$ is the approximate solution, $\tilde{\boldsymbol{\psi}}^n_\alpha$ is the vector of the nodal values of $\tilde{\psi}_\alpha^n$ and $\mathbf{N}(\mathbf{x})^T$ is the vector of piecewise-linear shape function centered at the grid nodes. By applying the Bubnov-Galerkin method, we finally obtain the discrete, weak form of Eq. \eqref{eq:local_formulation}
\begin{align}
\mathbf{M} (\tilde{\boldsymbol{\psi}}_\alpha^{n+1} - \tilde{\boldsymbol{\psi}}_\alpha^n) = \left(-\delta t \mathbf{C}_\alpha - \delta t^2 \mathbf{D}_\alpha \right) \tilde{\boldsymbol{\psi}}_\alpha^n,  \label{eq:galerkin_lbm}
\end{align}
where matrices $\mathbf{M}, \mathbf{C}_\alpha, \mathbf{D}_\alpha \in \mathbb{R}^{N_V \times N_V}$ are defined as
\begin{align}
\mathbf{M}        &= \int_{\mathcal{D}} \mathbf{N}\mathbf{N}^T dV \\
\mathbf{C}_\alpha &= \int_{\mathcal{D}} \mathbf{N} c_{\alpha r} \partial_r \mathbf{N}^T dV \\
\mathbf{D}_\alpha &= \frac{1}{2} \int_{\mathcal{D}} \partial_s\mathbf{N} c_{\alpha s}c_{\alpha r} \partial_r \mathbf{N}^T dV. \label{eq:fem_matrices}
\end{align}
In order to improve performance, instead of solving a linear system, we apply the lumped-mass approximation to matrix $\mathbf{M}$.

\subsection{Discrete Derivatives}
The remaining issue that needs to be addressed in the unstructured grid setting is the computation of discrete derivatives of the density $\rho$ and chemical potential $\mu_\varphi$ in Eqs. \eqref{eq:F}, \eqref{eq:G}, \eqref{eq:velocity} and \eqref{eq:pressure}. Since the density values are stored in grid nodes, they can be interpolated inside the elements using the linear shape functions as previously done in Eq. \eqref{eq:discrete_field}
\begin{align}
\rho({\mathbf{x}}) &= \mathbf{N}(\mathbf{x})^T \boldsymbol{\rho}.
\end{align}
$\mu_\varphi$ is treated the same way. The gradients of $\rho$ are then well-defined and constant inside each element
\begin{align}
\nabla \rho({\mathbf{x}}) &= \nabla\mathbf{N}(\mathbf{x})^T \boldsymbol{\rho}.
\end{align}
The nodal values of $\nabla \rho$ are then recovered using volume-weighted averaging of the element gradients
\begin{align}
\nabla\rho(\mathbf{x}_k) = \left(\sum_{T \ni \mathbf{x}_k} \mathcal{V}(T) \nabla\rho|_{\mathrm{int}(T)}\right) \bigg/ \sum_{T \ni \mathbf{x}_k} \mathcal{V}(T),
\end{align}
where $T$ iterates over all mesh elements containing the node $\mathbf{x}_k$, and $\mathcal{V}(T)$ denotes the volume of element $T$.

\section{Numerical Results}
We consider four benchmark flow problems to assess the validity and accuracy of the presented scheme. 

\subsection{Droplet in a Stationary Flow}
We first consider a droplet in a stationary flow. As this setup employs periodic boundary conditions, the physical properties of the model can be examined independently from the choice of boundary condition. 

\begin{figure}[h!]
\centering
\begin{minipage}{.475\textwidth}
  \centering
  \includegraphics[width=1.0\linewidth]{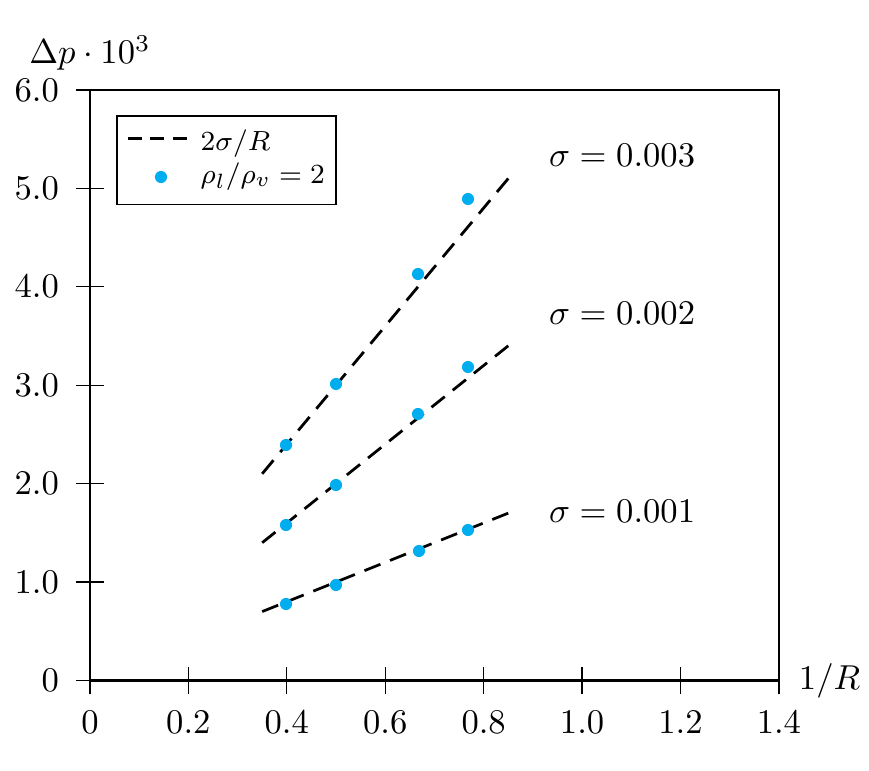}
  \captionof{figure}{Verification of Laplace's law on mesh $\mathcal{M}_2$.}
  \label{fig:laplace}
\end{minipage}\quad%
\begin{minipage}{.475\textwidth}
  \centering
  \includegraphics[width=1.0\linewidth]{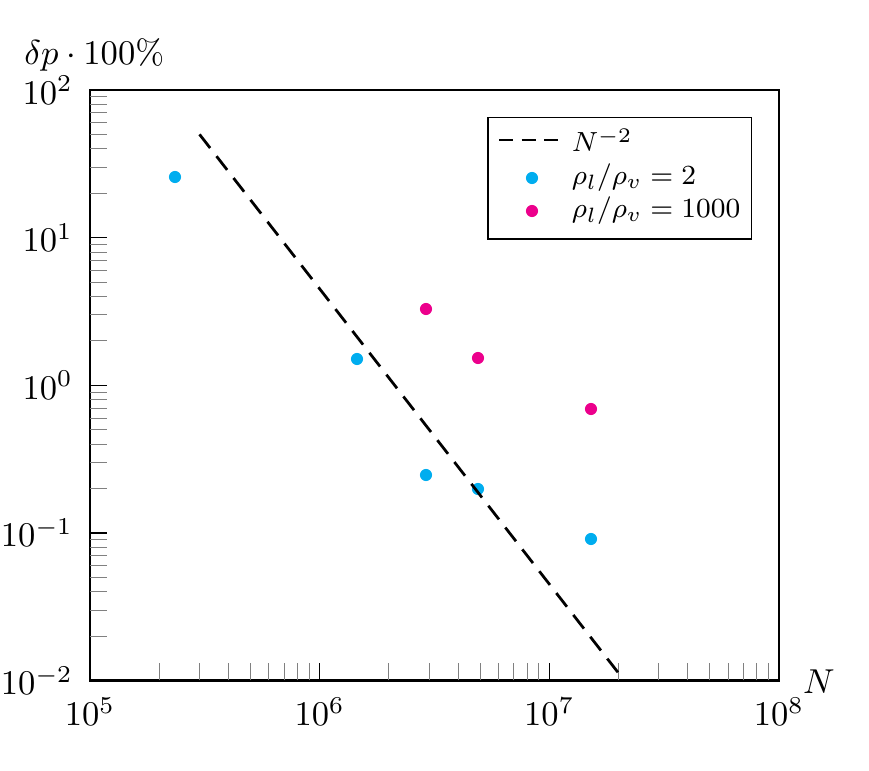}
  \captionof{figure}{Fractional error in pressure for meshes $\mathcal{M}_{1-5}$ in.}
  \label{fig:laplace_error}
\end{minipage}
\end{figure}

According to Laplace's law, the pressure difference across the interface of a three-dimensional droplet of radius $R$ at equilibrium is related to the surface tension via $p_{\text{in}}-p_{\text{out}} = 2\sigma/R$. We verify this relation by generating a droplet inside a cubic mesh and letting the system equilibrate. The pressure difference $\Delta p$ is measured by averaging the pressure inside $(R-\xi)$ and outside $(R+\xi)$ the droplet. The results are shown in Fig. \ref{fig:laplace} for three different values of surface tension and identical kinematic viscosities, displaying excellent agreement with theory. 

In Fig. \ref{fig:laplace_error} the fractional deviation $\delta p$ in pressure difference is illustrated for different grid sizes $N$ and density contrasts, indicating that the error approximately scales inversely to the square of the grid size. These results are summarized in Table \ref{tab:laplace_error}. 

\begin{table}[h!]
\centering
\caption{Fractional deviation in the simulated pressure difference $\Delta p$ relative the theoretical value $2\sigma/R$ with $\sigma=0.002$ and $R=2.5$ for different mesh resolutions.} 
\begin{tabular}{cccccc}
  Mesh &Elements $N$ &$\rho_l/\rho_v$ &$(\Delta p)\cdot 10^3$ &$\delta p\cdot 100\%$ \\
  \hline
  \hline				
  $\mathcal{M}_1$ &$2.4\cdot 10^5$    &2    &1.1902   &25.6119\%  \\ \\
  $\mathcal{M}_2$ &$1.5\cdot 10^6$    &2    &1.5762   &1.4906\%   \\ \\
  $\mathcal{M}_3$ &$2.9\cdot 10^6$    &2    &1.5961   &0.2456\%   \\
  --              &--                 &1000 &1.5477   &3.2694\%   \\ \\
  $\mathcal{M}_4$ &$4.9\cdot 10^6$    &2    &1.5969   &0.1969\%   \\
  --              &--                 &1000 &1.5757   &1.5194\%   \\ \\
  $\mathcal{M}_5$ &$1.5\cdot 10^7$    &2    &1.5986   &0.0900\%   \\
  --              &--                 &1000 &1.5891   &0.6794\%   \\  
  \hline
  \hline  
\end{tabular}
\label{tab:laplace_error}
\end{table}

\subsection{Diagonal Translation of a Droplet}
Similar to the hydrodynamic Galilean invariance test for multiphase flows, we now consider the motion of the droplet due to a constant velocity field $u_i = (u_0, 0, 0)$. The initially circular droplet of radius $L/4$ is placed in the middle of a periodic domain measuring $L \times L \times L$ with $L=6.28$ and the density and kinematic viscosity contrast is 1000 and 60, respectively.

\begin{figure}[h]
\centering
\begin{subfigure}{.4\textwidth}
  \centering
  \includegraphics[width=1.0\textwidth]{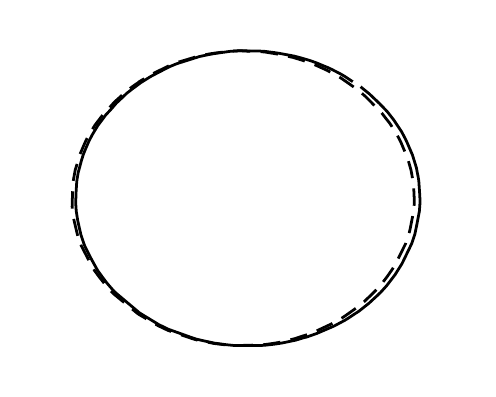}
  \caption{$\mathcal{M}_3$}
  \label{fig:droplet_shift3M}
\end{subfigure}%
\begin{subfigure}{.4\textwidth}
  \centering
  \includegraphics[width=1.0\textwidth]{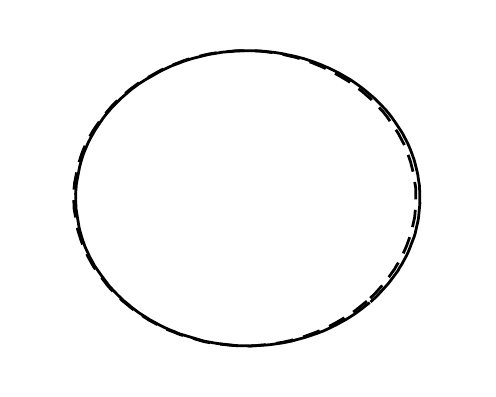}
  \caption{$\mathcal{M}_4$}
  \label{fig:droplet_shift5M}
\end{subfigure}
\caption{Profile of a droplet in a periodic domain with homogeneous velocity $u_i=(0.02, 0,0)$ at four round trips for two different mesh resolutions. The contour $\rho=(\rho_l + \rho_v)/2$ is shown: Dashed lines represent the translated profile and solid lines the initial profile.} 
\label{fig:droplet_shift}
\end{figure}

Fig. \ref{fig:droplet_shift} shows snapshots of the translated profile for the two meshes $\mathcal{M}_3$ and $\mathcal{M}_4$. In Fig. \ref{fig:droplet_shift3M} there is noticeable shift between the initial and final interface and the error in the eccentricity is 0.0036. The interfaces coincide well for the finer mesh and with an error of 0.0030 the relative error thus scales inversely to the square of the grid size as in the case of a static droplet (Fig. \ref{fig:laplace_error}). 

\subsection{Viscous Coupling in Concurrent Pipe Flow}
In complex, wall-bounded immiscible two-phase flows, the flow often aligns itself such the wetting phase flows along the solid surface, while the non-wetting phase flows in the center. In the current section we mimic this situation by investigating concurrent flow in a cylindrical pipe of radius $R$, where the non-wetting phase flows in the central region $r\in [0,a]$ and the wetting phase in the outer region $r\in (a,R]$. The steady-state radial velocity profile $u(r)$ follows from the Navier-Stokes equations, by utilizing the
symmetry of the problem
\begin{align}
u(r) = 
\begin{cases}
\frac{g}{4\mu_w}(R^2-r^2)                                 &\text{for $r\in (a,R]$}\\
\frac{g}{4\mu_w}(R^2-a^2) + \frac{g}{4\mu_{nw}}(a^2-r^2)  &\text{for $r\in [0,a]$}
\end{cases} \label{eqn:vel_cpf}
\end{align}
where $g$ is a constant volumetric body force that drives the flow along the axis, and $\mu_w$ $(\mu_{nw})$ denotes the dynamic viscosity of the wetting (non-wetting) phase. The dynamic viscosity contrast is denoted $M=\mu_{nw} / \mu _{w}$, which for the present case reduces to $M=\nu_{nw} / \nu _{w}$ as the densities are identical.

\begin{figure}[h!]
  \centering
  \includegraphics[width=0.6\linewidth]{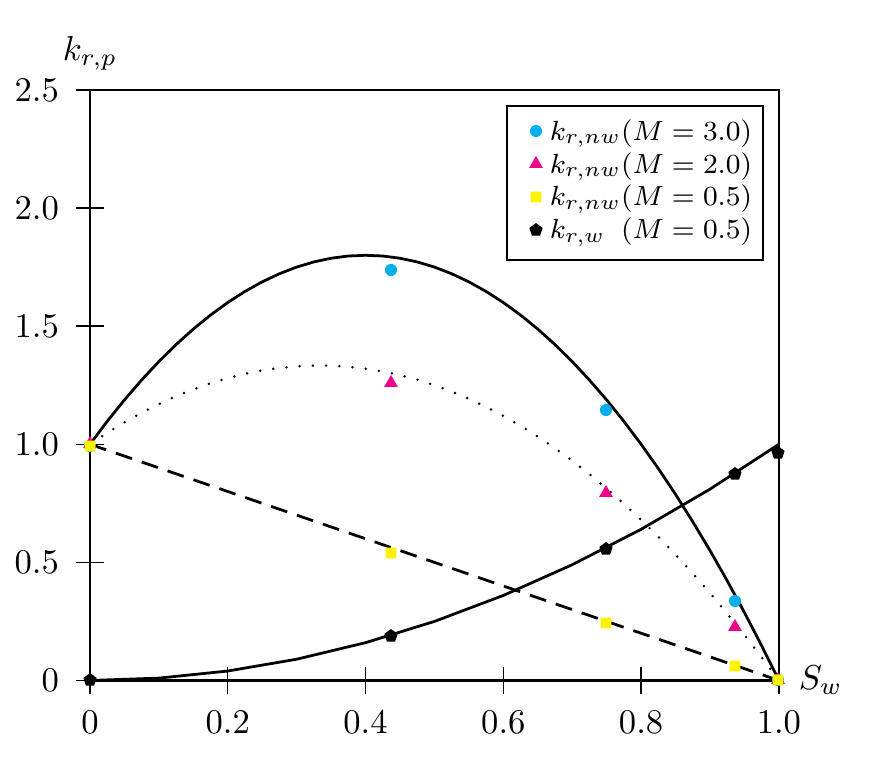}
  \caption{Relative permeability as a function of the wetting phase saturation for three different dynamics viscosity contrasts. The dotted, dashed and solid lines are the respective analytical expressions.}
  \label{fig::rel_perm_3D}
\end{figure}

The relative permeability $k_{r,p}$ of a phase $p\in \{w,nw \}$ is traditionally obtained as an adaption of Darcy's empirical law known from single-phase flows. As a function of the wetting saturation $S_w = 1-a^2/R^2$, $k_{r,p}$ is  defined in terms of the superficial Poiseuille flow rate $Q_p$ across a cross section perpendicular to the flow direction \cite{Yiotis200735}
\begin{align}
k_{r,p}(S_w) = \frac{1}{Q_p}\int_{p}\mathbf u_p \cdot d\mathbf{A}, \label{eqn:krp}
\end{align}
where the integration is performed over the phase. By combining \eqref{eqn:vel_cpf} and \eqref{eqn:krp} the analytical expressions for the relative permeabilities of the two phases in such system are given by
\begin{align}
k_{r,w}  &= S_w^2  \label{eqn:krw} \\
k_{r,nw} &= (1-S_w)[(1-S_w)-2((1-S_w)M-1)]. \label{eqn:krnw}
\end{align}
In Fig. \ref{fig::rel_perm_3D} $k_{r,p}$ is compared to the simulated values for different $M$, displaying good agreement for all wetting saturations. No-slip boundary condition is enforced using the bounce-back method at the solid boundary. The relative permeability of both phases is less than 1 for $M < 1$, as anticipated from \eqref{eqn:krw}-\eqref{eqn:krnw}. In contrast to the wetting phase, the relative permeability of the non-wetting phase depends on the viscosity contrast $M$, and for $M>1$ the flow simulation reveals that $k_{r,nw}$ is greater than the absolute permeability of the non-wetting phase for intermediate saturations. This well-known behavior is due to the lubricating effect of the wetting fluid on the non-wetting flow.

\color{black}
\subsection{Immiscible Rayleigh-Taylor Instability}
We now turn our attention to one of the most fundamental forms of interfacial instability between fluids of different densities, the Rayleigh-Taylor instability. The instability occurs when a perturbation is applied to the interface between a dense fluid on top of a lighter fluid in a gravitational field and has been studied by several methods thus far, see \emph{e.g.} \cite{PhysRevE.87.043301, Lee20131466, 1.869984, Wang201541}.

\subsubsection{Introduction}
Following the previous work by \cite{1.869984}, our system is confined to a three-dimensional rectangular box with height-width aspect ratio 4:1 and square horizontal cross-section. For simplicity the kinematic viscosity of the two fluids is chosen to be equal and surface tension is neglected. Periodic boundary conditions are applied at the four sides while no-slip boundary conditions are applied at the top and bottom walls. Gravity $g$ points downwards. 

The instability is developed from an initial single-mode perturbation $\epsilon$ with an amplitude 5\% of the domain width $\lambda$, 
\begin{align}
\epsilon(x,z)/\lambda = 0.05[\cos(2\pi x/\lambda) + \cos(2\pi z/\lambda)].
\end{align}
The characteristic parameters governing the flow are the Reynolds and Atwood number given by $\text{Re}=\lambda\sqrt{\lambda g}/\nu$ and $\text{At}=(\rho_l-\rho_g)/(\rho_l+\rho_g)$, respectively. In the following we present the results in dimensionless form, where $\lambda$ is taken as the length scale and $\sqrt{\lambda/g}$ the characteristic time scale. All simulations are performed on a mesh consisting of $1.1\cdot 10^7$ elements with $\text{Re}=256$ and $\text{At}=0.5$.

\begin{figure}[h]
\centering
\begin{subfigure}{.225\textwidth}
  \centering
  \includegraphics[width=1.0\textwidth]{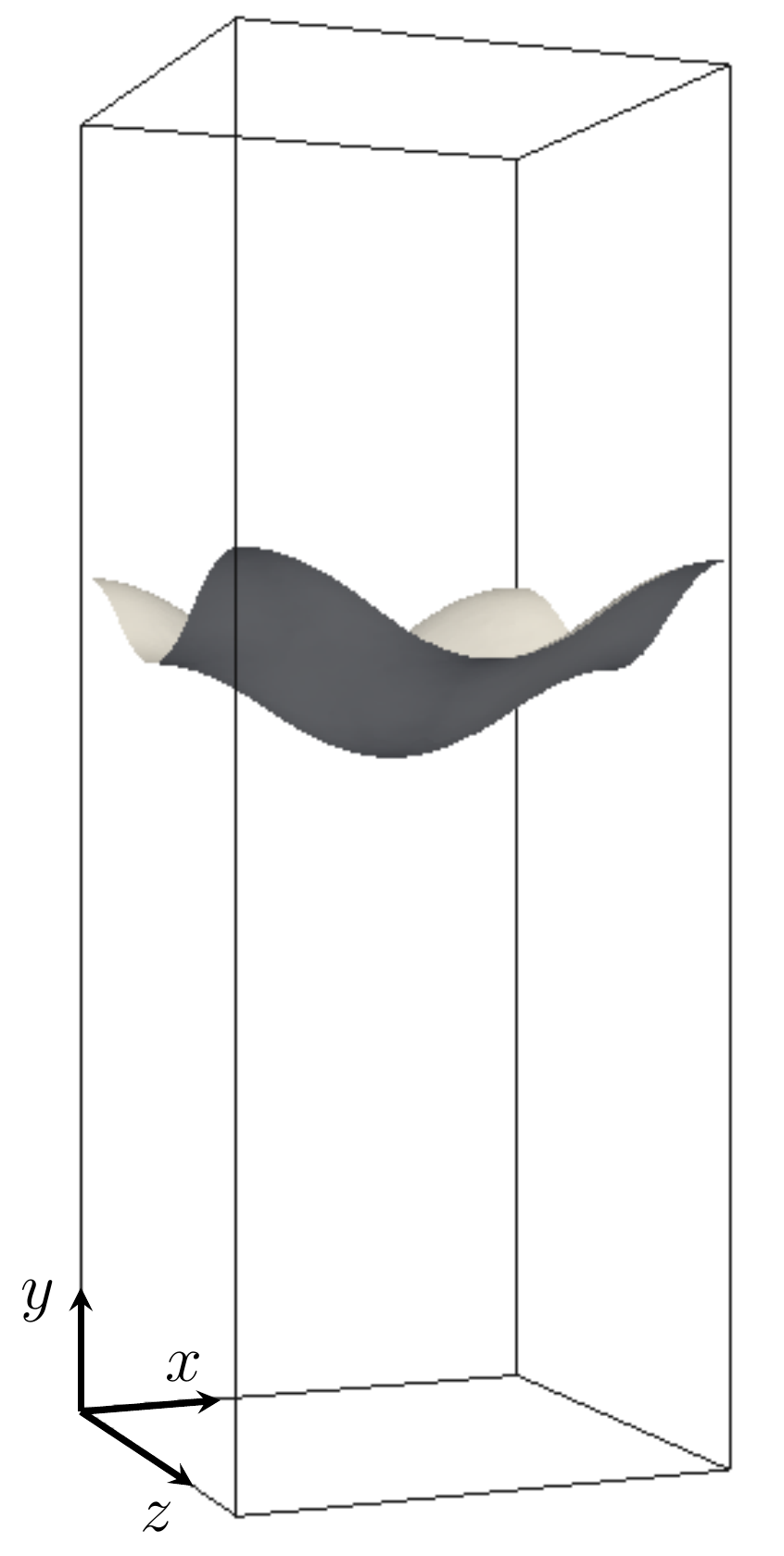}
  \caption{$t=1$}
  \label{fig:RT3D_1}
\end{subfigure}%
\begin{subfigure}{.225\textwidth}
  \centering
  \includegraphics[width=1.0\textwidth]{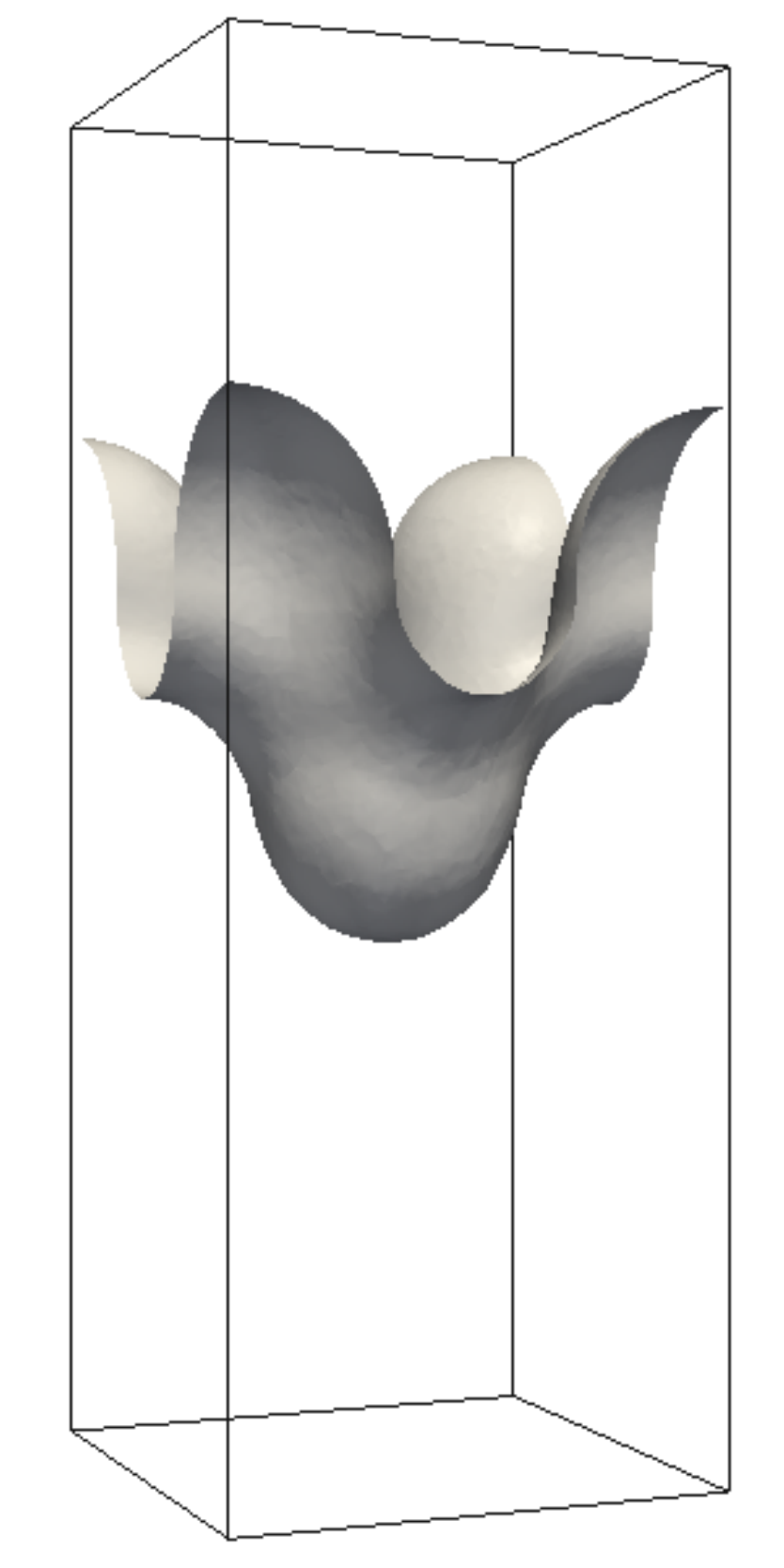}
  \caption{$t=2$}
  \label{fig:RT3D_2}
\end{subfigure}
\begin{subfigure}{.225\textwidth}
  \centering
  \includegraphics[width=1.0\textwidth]{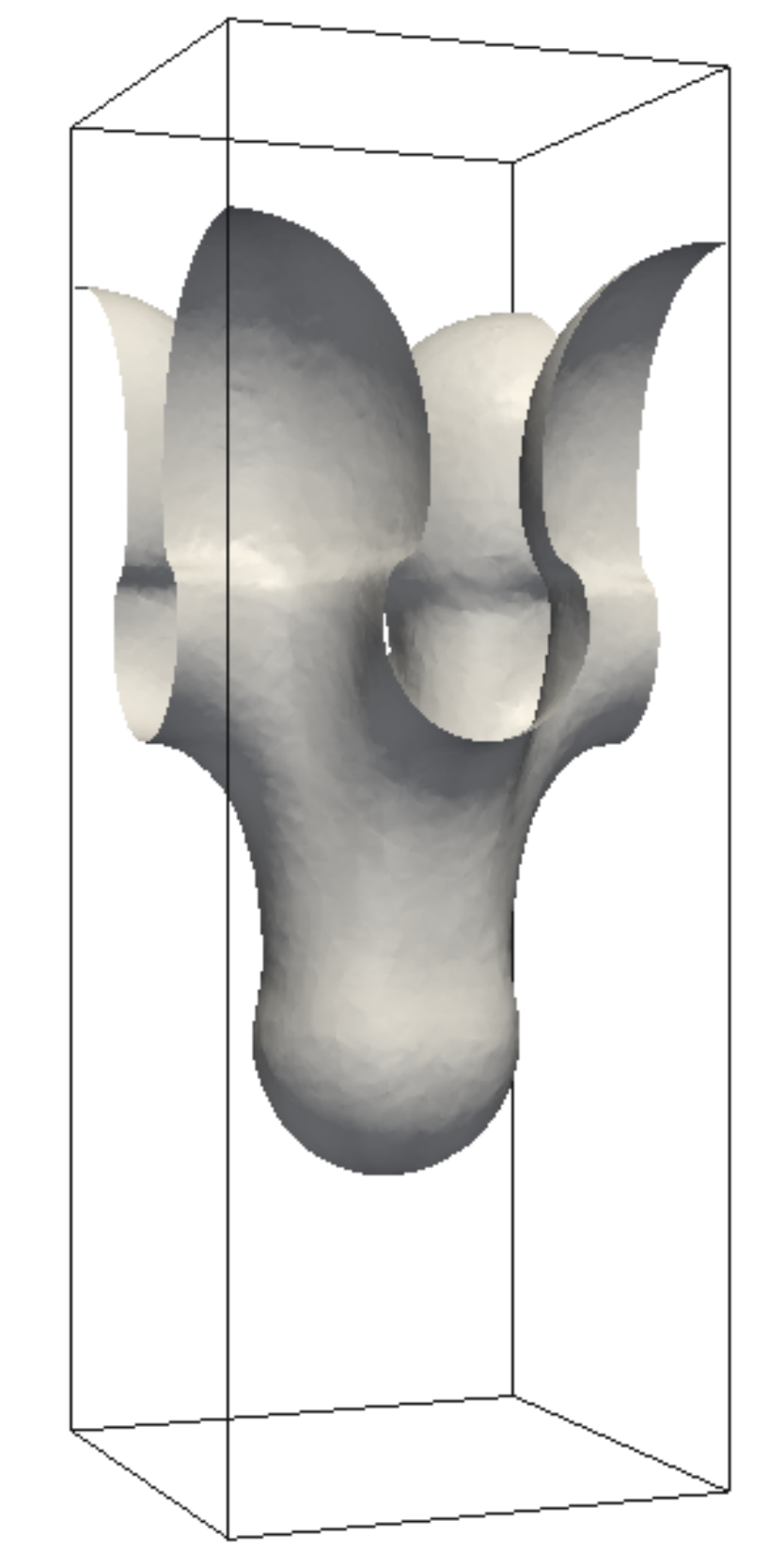}
  \caption{$t=3$}
  \label{fig:RT3D_3}
\end{subfigure}
\begin{subfigure}{.225\textwidth}
  \centering
  \includegraphics[width=1.0\textwidth]{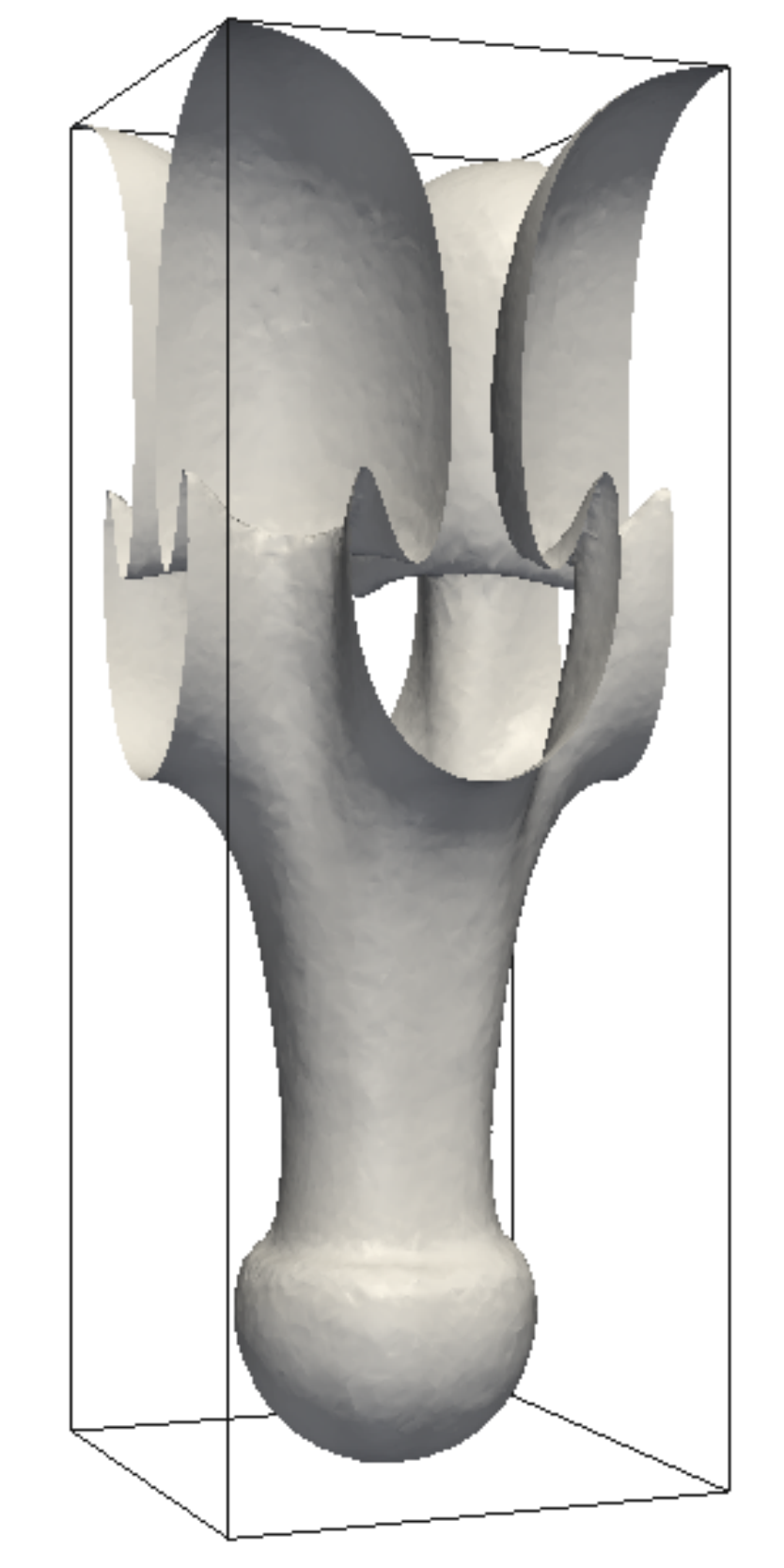}
  \caption{$t=4$}
  \label{fig:RT3D_4}
\end{subfigure}
\caption{Evolution of the interface in the three-dimensional Rayleigh-Taylor instability. The simulation is performed with a time step $\delta t \approx 34\tau$.}
\label{fig:RT3D}
\end{figure}

\subsubsection{Interface Dynamics}
\begin{figure}[h!]
\centering
\begin{subfigure}{.5\textwidth}
  \centering
  \includegraphics[width=0.6\textwidth]{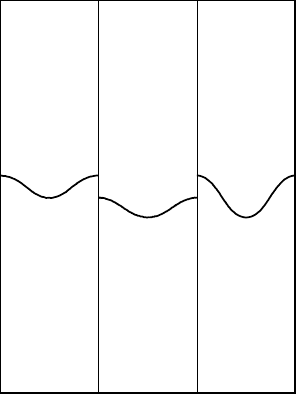}
  \caption{$t=1$}
  \label{fig:RT2D_1}
\end{subfigure}%
\begin{subfigure}{.5\textwidth}
  \centering
  \includegraphics[width=0.6\textwidth]{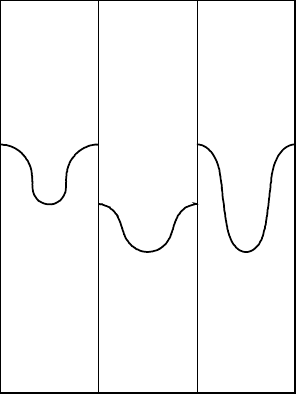}
  \caption{$t=2$}
  \label{fig:RT2D_2}
\end{subfigure}
\\%
\vspace{8mm}
\begin{subfigure}{.5\textwidth}
  \centering
  \includegraphics[width=0.6\textwidth]{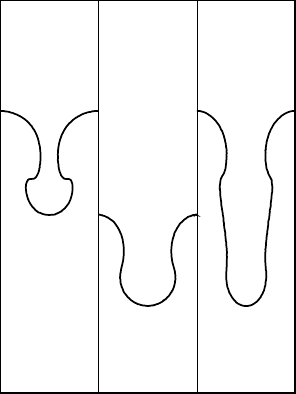}
  \caption{$t=3$}
  \label{fig:RT2D_3}
\end{subfigure}%
\begin{subfigure}{.5\textwidth}
  \centering
  \includegraphics[width=0.6\textwidth]{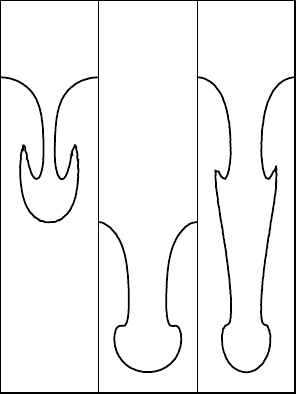}
  \caption{$t=4$}
  \label{fig:RT2D_4}
\end{subfigure}
\caption{Cross-sections of the interface at three vertical planes, $z=0$, $z=w/2$ and $z=x$.}
\label{fig:RT2D}
\end{figure}

The evolution of the interface is illustrated in Fig. \ref{fig:RT3D}. Initially, the interface grows symmetrically in the vertical direction at $t\simeq 1$ and remains simple. The evolution becomes more complicated with time, and at $t\simeq 2$ a single spike of the heavy fluid forms in the middle of the interface and bubbles of the light fluid rise along the periodic edges. As noted in \cite{1.869984}, a unique feature of the three-dimensional Rayleigh-Taylor instability is the emergence of saddle points in the middle of the four sides of the domain and the evolution around these. The first appearance of roll-ups of the dense fluid occurs in the neighbourhood of these saddle points, and at $t\simeq 3$ they have developed further. Roll-ups begin to form at the edge of the dense-fluid spike at the later time $t \simeq 4$, which eventually evolve into a mushroom-like shape. These observations are also apparent in the cross-sectional views displayed in Fig. \ref{fig:RT2D}. Only the interface along the diagonal plane $z=x$ differs from the two-dimensional problem, since this reveals the unique two-layer roll-up phenomenon.

\begin{figure}[h!]
  \centering
  \includegraphics[width=0.6\linewidth]{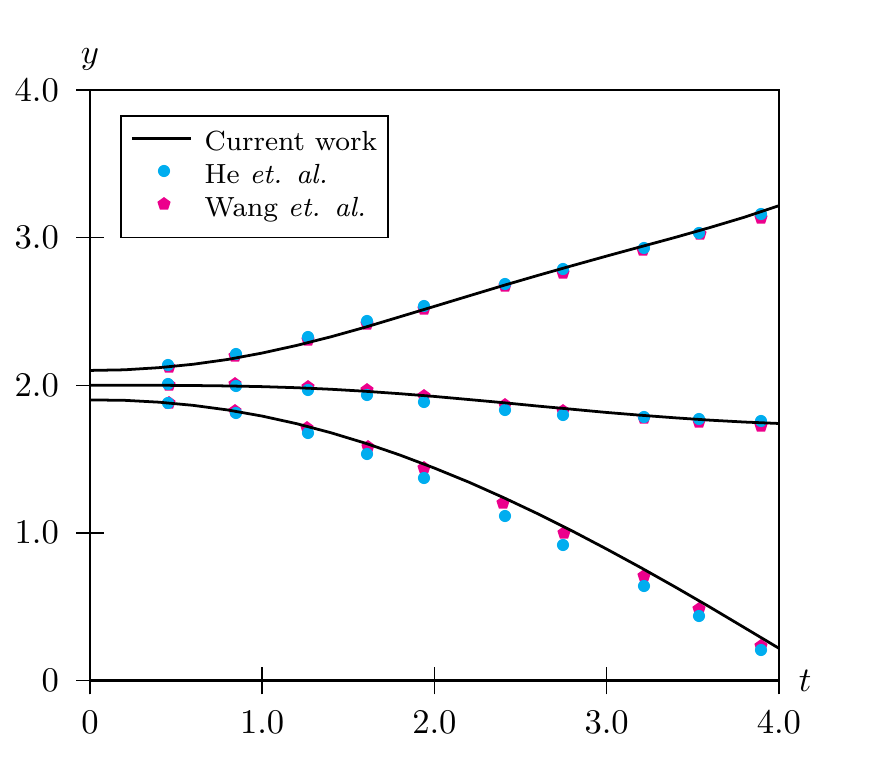}
  \caption{Temporal evolution of the bubble front (top), saddle point (middle) and spike tip (bottom) compared to \cite{1.869984, Wang201541}.}
  \label{fig::time_evolution}
\end{figure}

The trajectories of the light fluid bubble front, dense fluid spike tip and the saddle point are presented in Fig. \ref{fig::time_evolution}. The saddle point falls slowly during the entire evolution for this set of parameters, but the bubble and spike grow exponentially at early times, consistent with the theoretically expected growth. At later times, the bubble front grows with a constant velocity of $0.68$ in units of $\sqrt{0.5g\lambda\text{At}}$.



\section{Conclusion}
The presented scheme is based on a finite element lattice Boltzmann model and this choice, as opposed to approaches based on regular grids, is motivated by the higher flexibility and accuracy of irregular meshes at representing complex solid boundaries. The validity and grid convergence is established through simulations of benchmark problems that display excellent agreement with analytic results and literature data.
The ability of the model to simulate complex flows is verified through a study of the single-mode three-dimensional immiscible Rayleigh-Taylor instability.  

This work establishes a promising venue for simulations of multiphase flows in non-trivial geometries such as real porous media, \emph{e.g.}, by invoking the boundary treatment developed in \cite{Misztal2015316, 10.3389/fphy.2015.00050}.

\section*{Acknowledgement}
The authors acknowledge valuable discussions with Abbas Fakhari from the Department of Civil and Environmental Engineering and Earth Sciences, University of Notre Dame, Indiana 46556, USA. This work is financed by Innovation Fund Denmark and Maersk Oil and Gas A/S through the $\text{P}^3$ project.

\section*{References}
\bibliography{unstructlbm}

\end{document}